\documentclass[doublecol]{epl2}

\title{Averting group failures in collective-risk social dilemmas}

\shorttitle{Averting group failures in collective-risk social dilemmas}

\author{Xiaojie Chen,\inst{1}\footnote{E-mail: chenx@iiasa.ac.at} Attila Szolnoki,\inst{2}\footnote{E-mail: szolnoki.attila@ttk.mta.hu} Matja{\v z} Perc\inst{3}\footnote{E-mail: matjaz.perc@gmail.com}}
\shortauthor{Xiaojie Chen \etal}

\institute{\inst{1}Evolution and Ecology Program, International Institute for Applied Systems Analysis (IIASA), A-2361 Laxenburg, Austria\\
\inst{2}Institute of Technical Physics and Materials Science, Research Centre for Natural Sciences, Hungarian Academy of Sciences, P.O. Box 49, H-1525 Budapest, Hungary\\
\inst{3}Faculty of Natural Sciences and Mathematics, University of Maribor, Koro{\v s}ka  cesta 160, SI-2000 Maribor, Slovenia}


\abstract{\ Free-riding on a joint venture bears the risk of losing personal endowment as the group may fail to reach the collective target due to insufficient contributions. A collective-risk social dilemma emerges, which we here study in the realm of the spatial public goods game with group-performance-dependent risk levels. Instead of using an overall fixed value, we update the risk level in each group based on the difference between the actual contributions and the declared target. A single parameter interpolates between a step-like risk function and virtual irrelevance of the group's performance in averting the failure, thus bridging the two extremes constituting maximal and minimal feedback. We show that stronger feedback between group performance and risk level is in general more favorable for the successful evolution of public cooperation, yet only if the collective target to be reached is moderate. Paradoxically, if the goals are overambitious, intermediate feedback strengths yield optimal conditions for cooperation. This can be explained by the propagation of players that employ identical strategies but experience different individual success while trying to cope with the collective-risk social dilemma.}

\begin{document}
\maketitle

\section{Introduction}
Many of today's most pressing global challenges can be described as
``problems of the commons'' \cite{hardin_g_s68}. Be it the
preservation of natural resources for future generations, the
provisioning of health and social care, or the supply of energy to
meet our constantly increasing demand. All these challenges require
that we abandon some luxury on the personal level for the greater
good. The call goes out to all of us to intensify the level of
public cooperation across human societies \cite{ostrom_90}. However,
the temptations to free-ride on the efforts of others are strong,
especially since by nature we are hardwired to maximize our own
fitness regardless of the consequences this has for the public good.
Accordingly, the ``tragedy of the commons'' \cite{hardin_g_s68}
looms upon us, although we, the humans, are known for our highly
developed other-regarding abilities.

The public goods game is traditionally employed to study problems
that arise due to the dissonance between individual and societal
interests. During the game, all players that are members of a given
group have to decide simultaneously whether they wish to contribute
to the common pool or not. Regardless of their decision, each player
receives an equal share of the public good after the initial
contributions are multiplied by a synergy factor that is larger than
one. Evidently, individuals are best off by not contributing
anything, while the group is most successful if everybody
contributes. The competition between defection and cooperation has
received ample attention in the recent past, and several mechanisms
have been identified that promote prosocial outcomes. Examples
include voluntary participation
\cite{hauert_jtb02,hauert_s02,szabo_prl02}, inhomogeneous player
activities \cite{szolnoki_epl07,guan_pre07}, social diversity
\cite{perc_pre08,santos_n08,santos_jtb12}, appropriate partner
selection \cite{wu_t_epl09,zhang_hf_epl11}, aspiration-driven
mobility
\cite{helbing_pnas09,lin_yt_pa11,zhang_j_pa11,cardillo_axv12,yang_hx_pa12},
the introduction of punishment \cite{helbing_ploscb10,sigmund_n10,
helbing_njp10,rand_jtb09,szolnoki_pre11b,
isakov_dga12,sasaki_pnas12} and reward
\cite{szolnoki_epl10,hauert_jtb10}, coordinated investments
\cite{vukov_jtb11}, the Matthew \cite{perc_pre11} and joker effect \cite{requejo_pre12}, complex interaction networks
\cite{poncela_njp07,wang_z_epl12,poncela_njp09,tanimoto_pre12,vukov_pone11,
poncela_pre11,du_dga11}, conditional strategies \cite{szolnoki_pre12}, and
nonlinear benefit functions \cite{boza_bmceb10,archetti_ev11,deng_k_pone11b}, to name but a few
examples.

The public goods game in its classical form, however, fails to
capture some important features of social dilemmas that arise
frequently in realistic situations. A good example is the climate
change dilemma, where regions or nations may opt not to reduce their
carbon emissions in order to harvest short-term economic benefits.
Yet this is not the end of the story since failure to meet the
emission targets may have dire consequences in the future. The
so-called collective-risk social dilemma is more appropriate for
such a scenario \cite{milinski_pnas08}. There all players are
considered to have an initial endowment, and cooperation means
contributing a fraction of it to the common pool. Defectors do not
contribute. The risk level is determined by a collective target that
should be reached with individual investments. If a group fails to
reach this target, all members of the group loose their remaining
endowments with a certain probability.  Otherwise, everyone retains
its current endowment.
Experimental and theoretical studies have shown that high risks of collective failures raise the chances for coordinated actions \cite{milinski_pnas08,wang_j_pre09,greenwood_epl11,raihani_cc11},
and that this outcome is robust against variations of the interaction network and the size of the population \cite{santos_pnas11}.

In the pioneering works, the probability that endowments will be
lost or kept was most frequently considered to be a step-like
function of the gathered collective investments
\cite{milinski_pnas08,wang_j_pre09,santos_pnas11,greenwood_epl11}.
Hence, if the investments did not reach a certain fixed threshold
the probability to loose endowments was independent of the actual
contributions. Such a consideration, however, is not necessarily
accurate. It is intuitively easy to imagine cases where the
probability of a collective failure is much higher if the group
members are far from reaching the collective target, and vice versa
if the target is nearly yet not quite reached. Staying with the
climate change dilemma, it is reasonable to assume that the
escalation of problems is much more likely if the carbon emissions are
far in excess of the allowable quota than if they are just above it
\cite{raihani_cc11}.

Given these facts, we here propose that the risk level ought to
decreases continuously with increasing group investments, and we
investigate what are the consequences of the details of such an
upgrade on the evolution of cooperation in the public goods game
that is staged on a square lattice. In particular, we introduce a
function where a single parameter defines the feedback strength
between the actual performance of each group in relation to the
declared collective target and the risk level constituting the
probability that investments will be lost. While the larger the
difference between the target and the actual contributions the
higher the probability that all group members will loose their
investments, this dependence can be made more or less severe
depending on the feedback strength. By varying the latter and the
value of the collective target, we find that high targets require an
intermediate feedback strength for public cooperation to thrive,
while for moderate targets the higher the feedback strength the
better.

\section{Model}

\begin{figure}
\centering
\includegraphics[width=7cm]{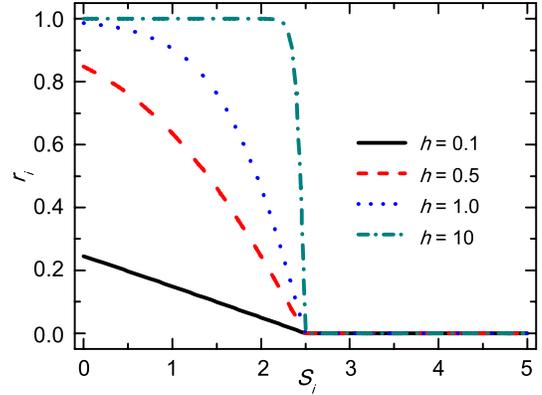}
\caption{Risk function in dependence on the group investment for different values of the feedback parameter $h$. For larger $h$ the traditionally considered step-like outlay is recovered. The collective target is $T=2.5$.} \label{fig1}
\end{figure}

\begin{figure}
\centering
\includegraphics[width=8cm]{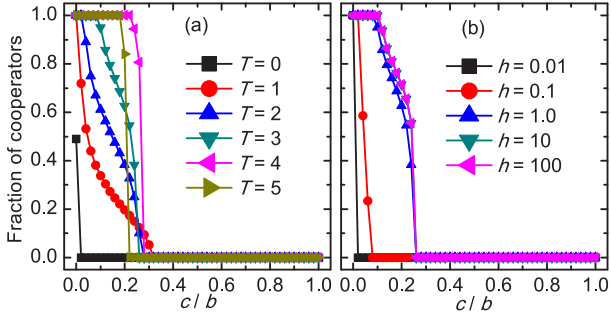}
\caption{Average cooperation level in dependence on the donation ratio $c/b$. (a) Using $h=0.5$ and different values of $T$. (b) Using $T=3$ and different values of $h$.} \label{fig2}
\end{figure}

As the interaction network, we consider a square lattice of size $L
\times L$ with periodic boundary conditions. Each player on site $x$
has an initial endowment $b=1$ and is designated as a cooperator
($s_x=1$) or defector ($s_x=0$) with equal probability. Cooperators
contribute an amount $c \leq b$ to the common pool while defector
contribute nothing. Moreover, there is a collective target $T$ to be
reached with the contributions in each group. If in group $G_i$ the
target is reached or surpassed, each member can keep its remaining
endowment. If not, all members loose their endowments with a
probability $r_i$, which is determined by a Fermi-type function
\begin{equation} r_i=\left\{
\begin{array}{lll}
tanh[(T-S_i)h] & \mbox{ if }S_i<T,\\
0 & \mbox{ if }S_i\geq T,
\end{array} \right.
\label{risk}
\end{equation}
where $h>0$ is the key parameter controlling the feedback strength
of failing to reach the target $T$ (see Fig.~\ref{fig1}), and
$S_i=\sum_{x\in G_i} s_x$ represents the total amount of collected
contributions in group $G_i$. Accordingly, player $x$ obtains its
payoff $P_x^i$ from group $G_i$.
Since players are connected to their four nearest neighbors,
each group has size $N=5$, and each player is member in five
overlapping groups, thus reaching a total payoff $P_x=\sum{_{i}}
P_x^i$.

After playing the game, each player $x$ is allowed to learn a
potentially better strategy from one of its randomly chosen
neighbors $y$ in agreement with the probability
\begin{equation}
f(P_y-P_x)=\frac{1}{1+\exp[(P_x-P_y)/\kappa]},
\end{equation}
where $\kappa$ denotes the amplitude of noise \cite{szabo_pr07}.
Without loss of generality we use $\kappa=0.5$, implying that better
performing player are readily adopted, but it is not impossible to
adopt the strategy of a player performing worse. The evolutionary
process is implemented with synchronous updating, where all players
first collect their payoffs and then alter their strategies
simultaneously.
Notably, before each round all the players obtain the endowment $b$.
To test the robustness of our findings, we have
verified that similar results are obtained by using asynchronous
updating.

Presented results were obtained on $100 \times 100$ sized lattices,
but remain intact also if a larger system size is used. The
cooperation level was determined as a key quantity according to
$L^{-2}\sum{_{x}}s_x(\infty)$, where $s_x(\infty)$ denotes the
strategy of player $x$ in the
stationary state where the average cooperation level becomes time-independent.

\section{Results}

\begin{figure}
\centering
\includegraphics[width=6.5cm]{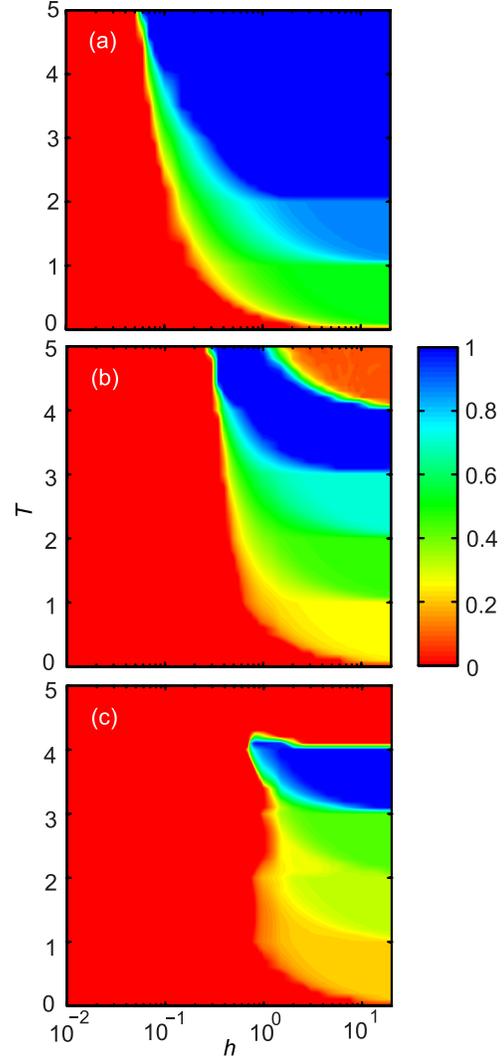}
\caption{Contours depicting the cooperation level in dependence on $h$ and $T$ for three different donation ratios: (a) $c/b=0.05$, (b) $c/b=0.2$, and (c) $c/b=0.25$.} \label{fig3}
\end{figure}

First, we show in Fig.~\ref{fig2}(a) the cooperation level
as a function of the donation ratio $c/b$ for five different values of
$T$ at a fixed intermediate feedback strength $h=0.5$. It can be
observed that the cooperation level decreases with increasing $c/b$
for all $T$. For $T=0$ our model behaves similarly to the
traditional public goods game in an unstructured population
\cite{archetti_ev11}. In this situation, the cooperation level is
zero for any $c/b>0$, while for $c/b=0$ it converges to $0.5$. When
$T$ is sufficiently high, full cooperation can be observed for
sufficiently small donation ratios. Interestingly however, for $T=5$
the performance is worse than for $T=4$; a detail that we will
elaborate on in what follows. Figure~\ref{fig2}(b) features
qualitatively similar results, only that the focus is on the impact
of $h$ at a fixed target $T=3$. It can be observed that larger
feedback strengths can sustain cooperation at larger $c/b$, although
the positive effect begins saturating for $h>1$.

In order to explore these effects more precisely, we present the
cooperation level in dependence on $h$ and $T$ together for three
representative values of $c/b$ in Fig.~\ref{fig3}. We find that
small values of $h$ (weak feedback) result in full defection for
each considered $c/b$ value and regardless of $T$. For intermediate
$h$, the cooperation level increases from zero to one upon
increasing $T$. For large $h$ (strong feedback), however, the
cooperation level first increases until a certain maximum is
reached, but then starts falling as $T$ increases further. If we
compare the cooperation level at a fixed target value, we find that,
in general, stronger feedbacks (higher $h$) yield better results.
Yet this is certainly not valid for high target values, where an
intermediate value of $h$ ensures much better conditions for the
evolution of public cooperation.

\begin{figure}
\centering
\includegraphics[width=7cm]{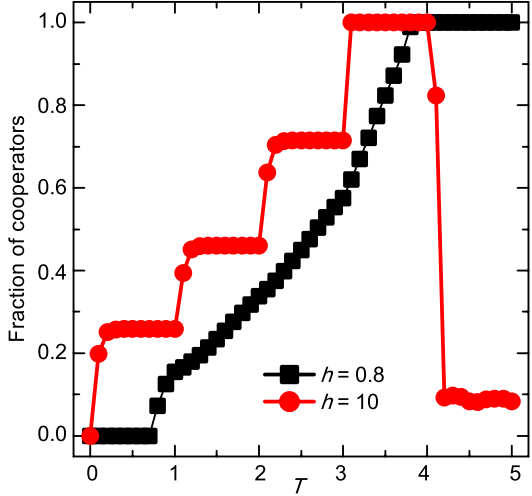}
\caption{Cooperation level in dependence on the collective target $T$ for two different values of $h$, both at $c/b=0.2$.} \label{fig4}
\end{figure}

This unexpected outcome is demonstrated separately in
Fig.~\ref{fig4}, where we plot the cooperation level as a function
of $T$ at two different values of $h$. As the figure shows, stronger
feedback generally results in a higher frequency of cooperators, but
this relation reverses at high values of $T$. The difference between
final states can be so large that applying intermediate $h$ yields a
full $C$ state, while for large $h$ the system arrives to a defector
dominated state.

To get an understanding of this rather paradoxical behavior, we
compare the time evolution of strategies at two representative
values of $T$ using the same $c/b=0.2$ donation ratio for three
different values of $h$. When plotting the spatial distribution of
strategies, it is useful to use different colors not just for
different strategies but also for the different levels of individual
success in terms of dealing with the collective-risk social dilemma.
More precisely, we distinguish players based on their ability to
collect payoffs from the majority of their groups or not.
Accordingly, a ``successful defector'' ($SD$, denoted yellow) is a
defector that can gather payoffs in at least three of the five
groups where it is involved. In the opposite case, the player is
marked as a ``failed defector'' ($FD$, denoted red). Identically, we
distinguish between ``successful cooperators'' ($SC$, denoted blue)
and ``failed cooperators'' ($FC$, denoted green).

\begin{figure}
\centering
\includegraphics[width=8cm]{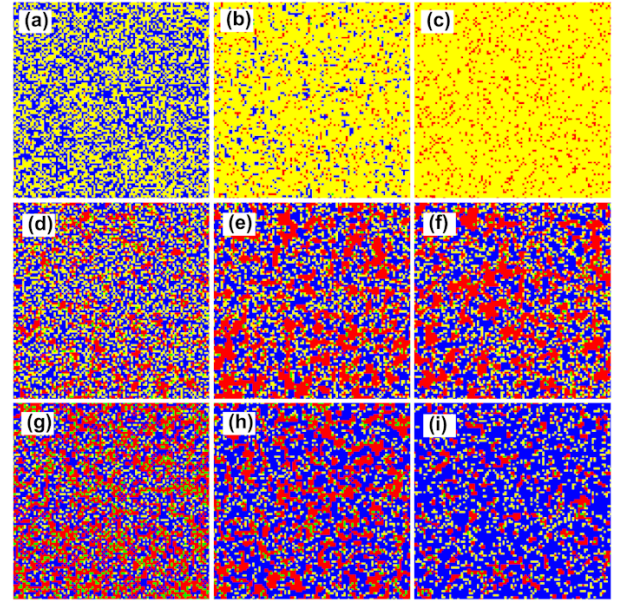}\\
\includegraphics[width=8cm]{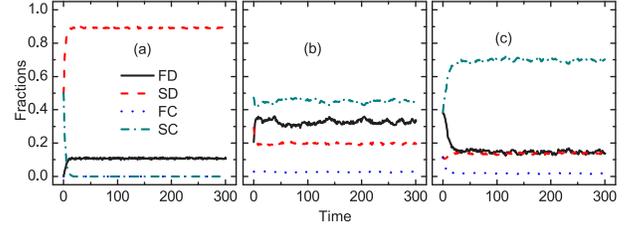}
\caption{Time evolution of sub-strategies as obtained for $h=0.1$ (top row), $h=0.8$ (middle row), and $h=10$ (bottom row), at $c/b=0.2$ and $T=2.6$ from left to right. Colors distinguish defectors who are able to gather payoffs in three or more groups (yellow) or at most in two groups (red). Cooperators are distinguished likewise and denoted blue if largely successful and green otherwise. Bottom panels depict the corresponding time evolutions of fractions of the four considered sub-strategies.} \label{fig5}
\end{figure}

Figure~\ref{fig5} shows the evolution from left to right at $T=2.6$
for $h=0.1$ (top row), $h=0.8$ (middle row), and $h=10$ (bottom
row). When the feedback is weak (top row) almost every player can
collect payoffs from the majority of the five groups where it is
member [$SD$ (yellow) and $SC$ (blue) players dominate in
Fig.~\ref{fig5}(a)]. Since the collective risk fails to avert from
antisocial behavior defectors can keep their benefit with a high
probability and cooperators therefore have no chance to survive. As
a result, the system terminates into a full $D$ state where there is
a dynamical balance between $SD$ and $FD$ players. Their spatial
distribution is uncorrelated, as shown in Fig.~\ref{fig5}(c), and
their density is directly related with the risk function, defined by
Eq.~\ref{risk}. Namely, the density of $FD$ players is proportional
to
\begin{equation}
\sum_{i=0}^{i < N/2} {N \choose i} (1-r_0)^i r_0 ^{N-i} \,.
\label{rho_FD}
\end{equation}

At larger $h$, shown in the middle row of Fig.~\ref{fig5}, the
impact of a higher collective risk becomes visible. Accordingly, the
number of $SD$ players decreases significantly and they can only
survive in the vicinity of cooperators. Because of notable
collective risks they cannot aggregate but need a spare distribution
to survive. The other type of defectors ($FD$, who failed to gather
payoffs in more than two groups) may form clusters, as can be
observed in Figs.~\ref{fig5}(e) and (f), because their state cannot
change rapidly. More precisely, while the transition from $SD \to
FD$ state may occur anywhere, even in the bulk of a defector island
the transition from $FD \to SD$ or from $FD \to SC$ state can only
happen via an imitation process at the interface of $FD$ domains.
This dynamical difference of transitions explains why $FD$ players
(red) are aggregated while $SD$ players (yellow) are distributed
homogeneously but close to $SC$ players (blue). It is also worth
mentioning that $FC$ players (green) occur rarely, typically in the
sea of failed defectors where the low density of cooperators cannot
warrant them to avoid the consequences of notable risk.

If using even larger $h$ values, as in the bottom row of
Fig.~\ref{fig5}, the above described mechanisms become even more
pronounced. Successful defectors are still able to utilize the
vicinity of cooperators to avoid the risk-dilemma, and hence their
density remains almost the same if compared to the smaller $h$
cases. This can be observed best from the bottom-most plots, which
depict the time evolution of the four sub-strategies [note that the
stationary fraction of $SD$ does not change significantly between
(b) and (c) panels]. The relevant change that lifts the fraction of
$FC$, and hence the cooperation level, is the shrinkage of $FD$
(red) islands. It is because the sharper risk probability makes the
invasion of $FC$ cooperators from the interface of $FD$ islands more
vigorous. According to this argument, it is generally clear why
increasing $h$ (stronger feedback) enhances the overall cooperation
level.

\begin{figure}
\centering
\includegraphics[width=8cm]{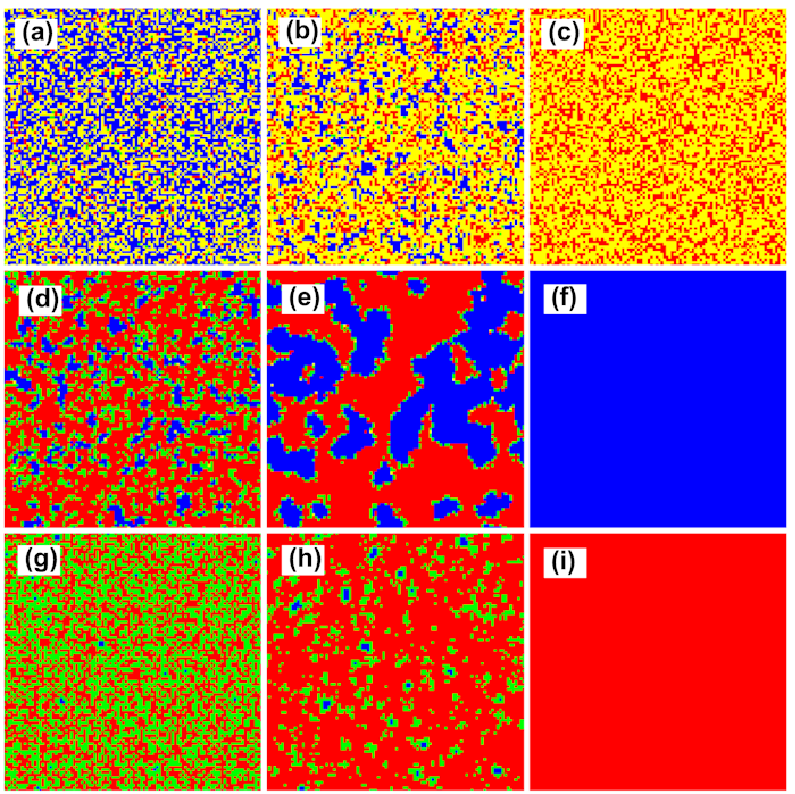}\\
\includegraphics[width=8cm]{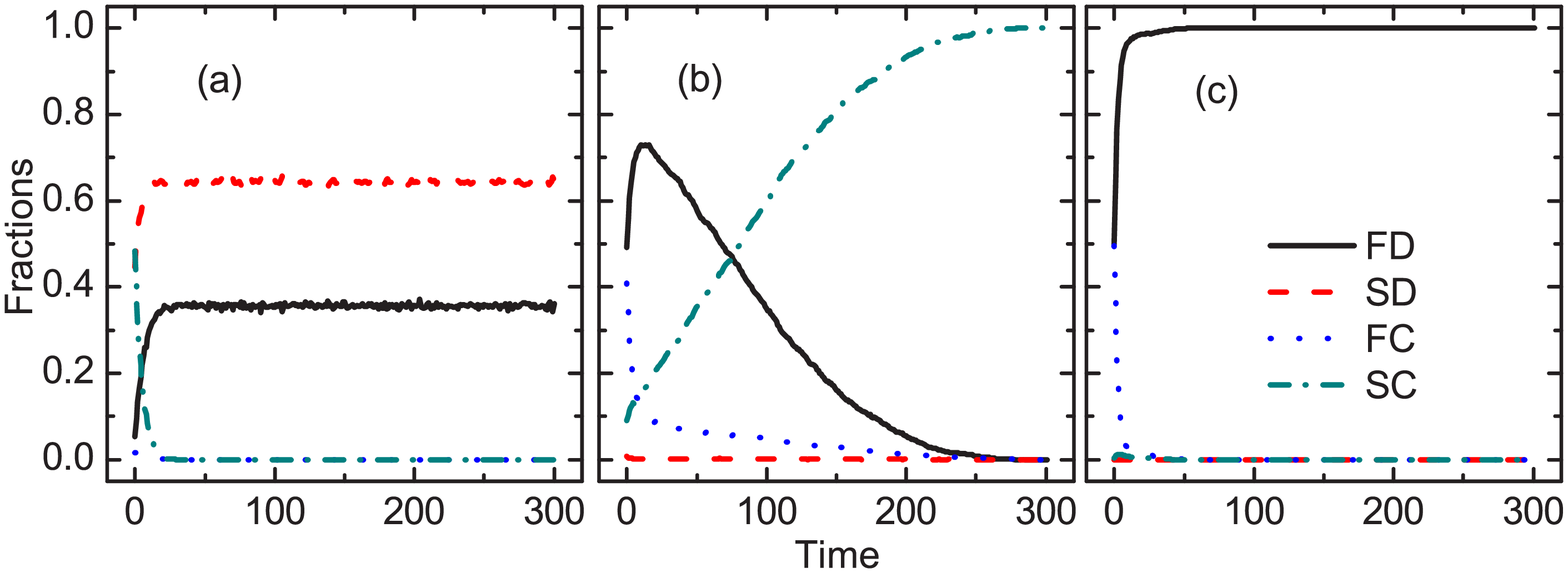}
\caption{Time evolution of sub-strategies as obtained for $h=0.1$ (top row), $h=0.8$ (middle row), and $h=10$ (bottom row), at $c/b=0.2$ and $T=4.5$ from left to right. The color scheme of sub-strategies is the same as in Fig.~\ref{fig5}. Bottom panels depict the corresponding time evolutions of fractions of the four considered sub-strategies.} \label{fig6}
\end{figure}

The above described mechanism is valid for almost all target values.
An important exception, however, are very high values of $T$, where
significantly different conclusion must be drawn. The unexpected
behavior is demonstrated in Fig.~\ref{fig6} where the same $c/b$
ratio and $h$ values were used as in Fig.~\ref{fig5}, but at
$T=4.5$. At small $h$ (top row of Fig.~\ref{fig6}), the players are
initially unsuccessful almost independently of their strategies.
This is because every group fails to fulfill the ambitious
collective target, which is simply too high. The success of one or
the other strategy is just the result of stochastic events driven by
the $r_i$ functions. In the later stages of the game defectors
eventually invade cooperators because the latter have to bare the
additional costs. Notably, cooperators cannot utilize the advantage
of clustering because of the smoothed $r_i$ function and the high
value of $T$. The final full $D$ state, plotted in
Fig.~\ref{fig6}(c), is similar to the one obtained for the smaller
$T$ value in Fig.~\ref{fig5}(c). The only difference is the higher
density of $FD$ players, which is due to the higher target and hence
the higher risk probability, which can again be estimated from
Eq.~\ref{rho_FD}.

At the intermediate $h$ value, as demonstrated in the middle row of
Fig.~\ref{fig6}, the significant change is that $SD$ players
disappear very soon, which is because the sharper outlay of the
$r_i$ function makes it unlikely for such defectors to avoid the
consequences of the now higher collective risk. $FD$ players can
spread temporarily because they avoid paying the cost, but later
they fail too, as illustrated by the black continuous curve in the
lowermost middle panel of Fig.~\ref{fig6}. Note that this is a
typical pattern that can be observed in spatial evolutionary games.
Because of the relatively unambiguous $r_i$ function, the support,
or lack thereof, of a group is clear. Hence, the islands of
cooperators become victorious. It is because they can always keep
their payoffs while defectors cannot. There are some failed
cooperators remaining, but they are predominantly restricted to the
frontiers of $SC$ domains. They are unlucky indeed, since of their
vicinity to defectors they have to share the sad consequences of
membership in a poorly (or at least insufficiently) productive
group. Paradoxically, they are the pioneers who begin invading
defective domains because they still have larger payoffs than
defectors. When a neighboring defector becomes cooperator the
mentioned $FC$ player may also transform to the $SC$ state with a
higher payoff. As a result, blue $SC$ domains invade red $FD$
islands and dominate the whole population. This invasion process and
the special role of $FC$ players is very similar to that of
conditional cooperators in a structured population, as shown very
recently in \cite{szolnoki_pre12}.

Even stronger feedbacks revert the described positive effect in the
opposite direction, as demonstrated in the bottom row of snapshots
in Fig.~\ref{fig6}. Here even an aggregation of cooperators is
mostly unable to fulfill the strict condition of reaching the high
collective target. Note that 12 cooperators should be accumulated
around the focal player for the latter to avoid the collective risk.
Even if this condition is met, there will be unsuccessful
cooperators (FC) at the edges of such domains that will be
vulnerable due to their inability to collect a similarly high payoff
and avoid the looming collective risk. Importantly, if using such a
sharp $r_i$ function it is irrelevant how close the group investment
is to the collective target: if the threshold is not met, the
``punishment'' will be the same as in a fully defective group.
Consequently, unlike in the intermediate $h$ case, unsuccessful
cooperators cannot invade defectors, which ultimately results in
complete defector dominance, as depicted in Fig.~\ref{fig6}(i).

\section{Summary}
We have studied the collective-risk social dilemma in a structured
population, focusing on the emergence of public cooperation under
the influence of differently shaped risk functions. Most
importantly, we have considered the risk level to decreases
continuously with increasing group investments, with a single
parameter enabling us to interpolate between different feedback
strengths of the difference with regards to the declared collective
target. In agreement with previous observations, we have shown that
sharper risk functions, corresponding to a stronger feedback, in
general promote the evolution of public cooperation and may thus
help to prevent the tragedy of the commons. Yet we have found this
to hold only if the collective targets are sufficiently moderate. If
the goals in terms of the production of public goods are too high,
intermediate feedback strengths can yield much higher levels of
public cooperation than strong feedbacks. This goes against
preliminary expectations, signaling that the expectation for most of
the group members to contribute maximally to the common pool is a
difficult proposition that requires a special approach. It is
certainly not impossible to achieve, but requires a certain degree
of lenience towards all that are involved. An overall high risk of
collective failure is then certainly not advisable, but rather one
should consider diverse and fine-grained risk intervals that are
able to take into account how far away the production of any given
group is from the declared target. We have revealed key mechanisms
that are responsible for these observations by introducing
sub-strategies that further divide the traditional cooperators and
defectors based on their individual success in groups they are
involved with, thus complementing previous studies
\cite{wang_j_pre09,santos_pnas11} and hopefully promoting our
understanding of the evolution of public cooperation in the
collective-risk social dilemma.

\acknowledgments
Financial support from the Hungarian National Research Fund (grant K-101490) and the Slovenian Research Agency (grant J1-4055) is gratefully acknowledged.

\end{document}